\documentclass[journal=ancac3,manuscript=article]{achemso}

\usepackage{chemformula} 
\usepackage[T1]{fontenc} 

\author{Evgeny M. Alexeev}
\email{ea529@cam.ac.uk}
\affiliation[University of Sheffield]
{Department of Physics and Astronomy, University of Sheffield, Sheffield S3 7RH, UK}
\author{Nic Mullin}
\affiliation[University of Sheffield]
{Department of Physics and Astronomy, University of Sheffield, Sheffield S3 7RH, UK}
\author{Pablo Ares}
\affiliation[University of Manchester]
{Department of Physics and Astronomy, University of Manchester, Oxford Road, Manchester M13 9PL, UK}
\alsoaffiliation[National Graphene Institute]{National Graphene Institute, University of Manchester, Booth Street East, Manchester M13 9PL, UK}
\author{Harriet Nevison-Andrews}
\affiliation[University of Manchester]
{Department of Physics and Astronomy, University of Manchester, Oxford Road, Manchester M13 9PL, UK}
\alsoaffiliation[National Graphene Institute]{National Graphene Institute, University of Manchester, Booth Street East, Manchester M13 9PL, UK}
\author{Oleksandr V. Skrypka}
\author{Tillmann Godde}
\affiliation[University of Sheffield]
{Department of Physics and Astronomy, University of Sheffield, Sheffield S3 7RH, UK}
\author{Aleksey Kozikov}
\affiliation[University of Manchester]
{Department of Physics and Astronomy, University of Manchester, Oxford Road, Manchester M13 9PL, UK}
\alsoaffiliation[National Graphene Institute]{National Graphene Institute, University of Manchester, Booth Street East, Manchester M13 9PL, UK}
\author{Lee Hague}
\affiliation[University of Manchester]
{Department of Physics and Astronomy, University of Manchester, Oxford Road, Manchester M13 9PL, UK}
\alsoaffiliation[National Graphene Institute]{National Graphene Institute, University of Manchester, Booth Street East, Manchester M13 9PL, UK}
\author{Yibo Wang}
\affiliation[University of Manchester]
{Department of Physics and Astronomy, University of Manchester, Oxford Road, Manchester M13 9PL, UK}
\alsoaffiliation[National Graphene Institute]{National Graphene Institute, University of Manchester, Booth Street East, Manchester M13 9PL, UK}
\author{Kostya S. Novoselov}
\affiliation[University of Manchester]
{Department of Physics and Astronomy, University of Manchester, Oxford Road, Manchester M13 9PL, UK}
\alsoaffiliation[National Graphene Institute]{National Graphene Institute, University of Manchester, Booth Street East, Manchester M13 9PL, UK}
\author{Laura Fumagalli}
\affiliation[University of Manchester]
{Department of Physics and Astronomy, University of Manchester, Oxford Road, Manchester M13 9PL, UK}
\alsoaffiliation[National Graphene Institute]{National Graphene Institute, University of Manchester, Booth Street East, Manchester M13 9PL, UK}
\author{Jamie K. Hobbs}
\author{Alexander I. Tartakovskii}
\email{a.tartakovskii@sheffield.ac.uk}
\affiliation[University of Sheffield]
{Department of Physics and Astronomy, University of Sheffield, Sheffield S3 7RH, UK}

\title{Emergence of highly linearly polarized interlayer exciton emission in MoSe$_2$/WSe$_2$ heterobilayers with transfer-induced layer corrugation}
\begin{document}

\begin{tocentry}
\includegraphics{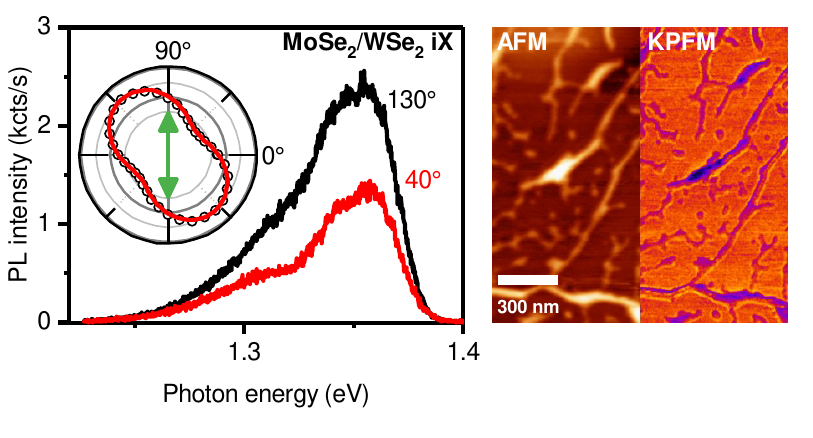}
\end{tocentry}

\begin{abstract}
The availability of accessible fabrication methods based on deterministic transfer of atomically thin crystals has been essential for the rapid expansion of research into van der Waals heterostructures. An inherent issue of these techniques is the deformation of the polymer carrier film during the transfer, which can lead to highly non-uniform strain induced in the transferred two-dimensional material. Here, using a combination of optical spectroscopy, atomic force and Kelvin probe force microscopy, we show that the presence of nanometer scale wrinkles formed due to transfer-induced stress relaxation can lead to strong changes in the optical properties of MoSe$_2$/WSe$_2$ heterostructures and the emergence of the linearly polarized interlayer exciton photoluminescence. We attribute these changes to the local breaking of crystal symmetry in the nanowrinkles, which act as efficient accumulation centers for the interlayer excitons due to the strain-induced interlayer band gap reduction. The surface potential images of the rippled heterobilayer samples acquired using Kelvin probe force microscopy reveal the variation of the local work function consistent with the strain-induced band gap modulation, while the potential offset observed at the ridges of the wrinkles shows a clear correlation with the value of the tensile strain estimated from the wrinkle geometry. Our findings highlight the important role of the residual strain in defining optical properties of van der Waals heterostructures and suggest novel approaches for interlayer exciton manipulation by local strain engineering.
\end{abstract}

\newpage
\subsection{Introduction}
\begin{figure}[ht]
	\centering
	\includegraphics{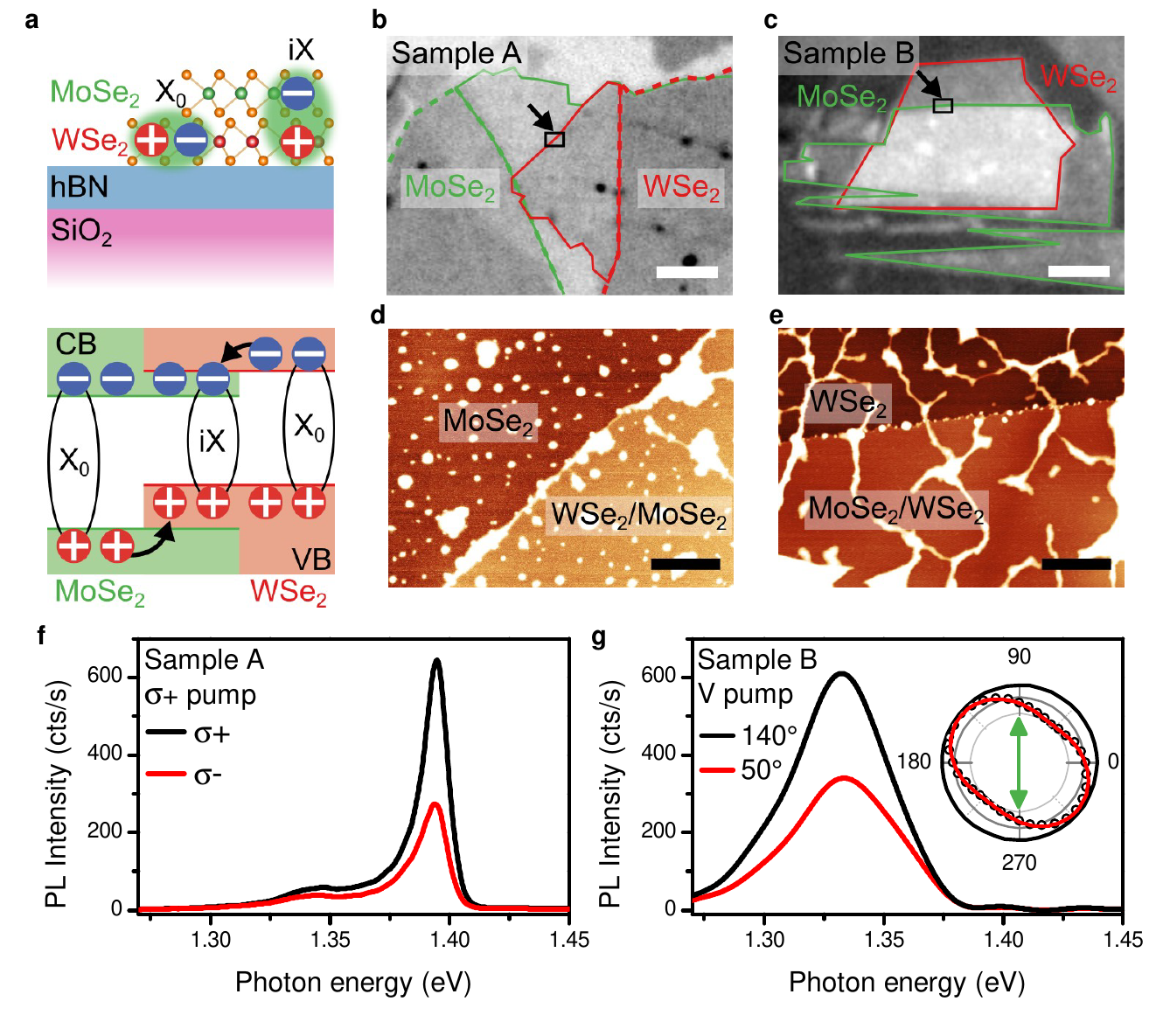}
	\caption{\textbf{Interlayer exciton photoluminescence in samples with different layer morphology.} (a) Schematic representation of device structure (top) and band alignment (bottom) in MoSe$_2$/WSe$_2$ samples, showing the formation of the intralayer (X$_0$) and interlayer (iX) excitons by carriers confined within the same or different layers, respectively (b and c) Optical images of MoSe$_2$/WSe$_2$ samples A and B assembled on hBN/SiO$_2$/Si substrates by deterministic dry-peel transfer. Edges of the monolayer TMD regions are shown by solid lines. Contrast of the digital images was artificially increased to make the monolayer regions more visible. Scale bars, 3 $\mu$m (d and e) Topographic AFM images recorded in the regions of samples A and B indicated by black rectangles in the optical images. Transfer-induced nanowrinkles are clearly visible in sample B. Black to white scale, 4 nm. Scale bars, 200 nm. (f and g) Low-temperature photoluminescence spectra recorded in samples A (f) and B (g), centered at the interlayer exciton range. While sample A demonstrates the retention of incident circular polarization, indicating the generation of strong iX valley polarization, the emission in sample B is highly linearly polarized, with the polarization axis rotated by 50$^{\circ}$ with respect to the excitation laser polarization. Inset plots integrated PL intensity in sample B as a function of the detection angle; incident laser polarization direction is marked by the green arrow.}
	\label{fig:SampleImage}
\end{figure}

Two-dimensional (2D) semiconductors, such as monolayer transition metal dichalcogenides (TMDs), have emerged as a promising platform for photonic and optoelectronic applications due to their strong light-matter interactions\cite{Zhang2014,Li2014b,Chernikov2014,Epstein2019}, direct band gap in visible-to-near-infrared frequency range\cite{Wang2012,Liu2015a,Manzeli2017}, and unique valley-contrasting physical effects\cite{Xu2014,Schaibley2016,Xiao2018,Binder2019}. 
Van der Waals (vdW) heterostructures assembled from TMD monolayers have attracted particular research interest due to their ability to host interlayer excitons (iXs) – bound complexes of electrons and holes located in adjacent layers (Fig.~\ref{fig:SampleImage}a)\cite{Baranowski2017a,Okada2018,Rivera2018,Wang2019b}. 
Similar to their intralayer counterparts, iXs have binding energies of a few hundred meV\cite{Wilson2016,Ovesen2018a} and exhibit valley-dependent optical selection rules, allowing selective excitation and probing of valley-polarized exciton population using circularly-polarized light\cite{Rivera2016}. However, they show orders of magnitude longer radiative and valley polarization lifetimes due to the reduced spatial overlap of the electron-hole wave functions\cite{Rivera2015, Miller2017,Jiang2018a}. The properties of iXs can be further controlled by changing the rotational alignment between the layers\cite{Nayak2017,Kunstmann2018a,Alexeev2019,Seyler2019}. In particular, the emergence of long-range periodic modulation of the local crystal structure known as the moir\'e pattern can considerably alter the energy spectrum of excitons in the heterostructure\cite{Tran2019a, Jin2018c}, leading to the formation of topological\cite{Tong2016,mcdonald_intra,Chittari2019} and quantum-confined states\cite{hongyi_moire,Pan2018a,Brotons-Gisbert2019}.

Despite considerable progress in heterostructure fabrication through epitaxial growth\cite{Tan2018,Chen2018a,Cai2018}, mechanical stacking of layers remains the method of choice for research\cite{Zhou2018,Lv2019}. This methods includes two key steps: preparation of 2D crystals through mechanical exfoliation or chemical vapour deposition, and transfer of these crystals onto a target substrate at a selected location, usually containing another monolayer flake or a stack of atomically thin layers. This accessible and reliable technique provides unprecedented control over the layer order, composition, and thickness, allowing the creation of artificial materials with tailor-made properties and their subsequent integration into photonic\cite{Mak2016}, plasmonic\cite{Tran2017a,Luo2018} and superconducting\cite{Lee2018a} structures through deterministic placement. The ability to combine materials with different crystal symmetries and lattice parameters, as well as to control the interlayer rotation, makes this method essential for exploring these unique degrees of freedom. A variety of methods for mechanical transfer of few-layer materials has been developed, but the common feature that they share is the use of a flexible polymer film as a transfer medium\cite{Frisenda2017a}. Stretching and compression of the carrier film during pick-up or deposition can lead to delamination and buckling of the transferred material, resulting in a high level of residual strain in the sample. The formation of transfer-induced wrinkles has been observed for transfer methods based on polymethyl methacrylate (PMMA)\cite{Uwanno2015}, polypropylene carbonate (PPC)\cite{Pizzocchero2016}, and polydimethylsiloxane (PDMS)\cite{Castellanos-Gomez2014,Jain2018} carrier films. 

Here we show that the strong and highly anisotropic strain induced in two-dimensional layers during fabrication can substantially alter the optical properties of vdW heterostructures. Using MoSe$_2$/WSe$_2$ heterostructures assembled by PMMA-assisted dry-peel transfer as a case study, we demonstrate that the presence of residual strain concentrated in transfer-induced wrinkles can result in a substantial change in the properties of the interlayer excitons and the emergence of their linearly-polarized photoluminescence. The typical nanoscale dimensions of these features prevent their observation in optical microscopy images of the samples, requiring the use of atomic force microscopy (AFM) for their detection. We attribute the observed changes of the optical properties to local symmetry breaking at the nanowrinkles which act as trapping centers for interlayer exciton due to strain-induced interlayer band gap reduction. We investigate the potential landscape of the wrinkled TMD samples using Kelvin probe force microscopy (KPFM) and show that, while flat regions of the heterostructure demonstrate a homogeneous surface potential, a strong variation of the local work function occurs at the nanowrinkles, consistent with strain-induced band modulation\cite{Roldan2015}. Our findings highlight the important role of strain variations in defining the optical properties of vdW heterostructures and suggest novel approaches for controlled manipulation of iX using highly-localized strain fields. 

Figure~\ref{fig:SampleImage} b and c show bright-field microscope images of two MoSe$_2$/WSe$_2$ heterostructures fabricated by PMMA-assisted dry-peel transfer (see Methods for a full description of the fabrication procedure). The contrast of the images was digitally enhanced to make the TMD layers more visible. The heterostructures were fabricated by consecutively transferring monolayers of MoSe$_2$ and WSe$_2$ from the polymer carrier film onto thin hBN flakes that were mechanically exfoliated onto a SiO$_2$/Si substrate. While the two samples have a similar appearance in the optical images, the AFM images shown in Figure~\ref{fig:SampleImage}d and e reveal a striking difference in their surface morphology. Fig.~\ref{fig:SampleImage}d shows an AFM image of sample A, centered at the edge of the MoSe$_2$ monolayer and the WSe$_2$/MoSe$_2$ heterobilayer regions (area marked by the black rectangle in Fig.~\ref{fig:SampleImage}b). Despite a number of polymer residue particles visible in the image as white spots of various shapes, the surface does not show any visible signs of layer distortion, indicating that the contaminants are present at the exterior of the sample and not trapped between the layers\cite{Khestanova2016}. 

In contrast, while the surface of sample B (Fig.~\ref{fig:SampleImage}e) has an overall much cleaner appearance, notable layer corrugation is visible in both WSe$_2$ monolayer and MoSe$_2$/WSe$_2$ heterostructure regions. The wrinkles have a typical width of 15-30~nm at the base and a height of 1-3~nm, making them invisible in the optical images. The spatial distribution of these features indicates that they originate from the WSe$_2$ layer that was deposited onto the hBN in the first transfer step: they appear across the entire area of the WSe$_2$ flake and extend into the MoSe$_2$/WSe$_2$ heterostructure region, but are absent in the underlying hBN, as well as the in MoSe$_2$ layers deposited directly onto the hBN (see SI Fig.~S1). 

Anisotropic strain relaxation through the formation of nanowrinkles leads to a strong modification of the WSe$_2$ PL spectrum, resulting in substantial broadening of the emission peaks and the emergence of a strong localized emitter band (SI~Fig.~S2). It also leads to very pronounced changes in the properties of the iXs in the heterostructure region. Fig.~\ref{fig:SampleImage}f and g compare iX PL spectra recorded in the two samples at T~=~10~K (see Methods for optical experiment description). Sample A (Fig.~\ref{fig:SampleImage}f) with a flat surface morphology shows a PL spectrum consistent with typical iX emission observed in high-quality MoSe$_2$/WSe$_2$ samples reported in the literature\cite{Ciarrocchi2018, Hanbicki2018}: the iX PL peak is composed of two spectrally narrow components centered at 1.34 and 1.39 eV. Polarization-resolved PL spectra recorded under circularly-polarized excitation show that the emission co-polarized with the excitation has a higher intensity, indicating that valley polarization of both monolayers induced by the optical pumping is preserved by the iX during its lifetime\cite{Rivera2016}. The polarization degree of the higher-energy component, defined as $\rho = (I_{co} - I_{cr})/(I_{co} + I_{cr})$, where $I_{co}$ and $I_{cr}$ are the intensities of co- and cross-polarized emission, reaches a maximum value of 0.5 when the excitation light energy is nearly-resonant with the WSe$_2$ trion transition (1.7 eV). It is, however, strongly reduced for non-resonant excitation, dropping by almost an order of magnitude ($\rho \approx 0.05$) when the excitation energy is increased to 1.88 eV (see SI~Fig.~S3), consistent with earlier literature reports\cite{Hanbicki2018}.

Conversely, the PL spectrum recorded in sample B (Fig.~\ref{fig:SampleImage}g) shows a much broader iX peak with the intensity maximum located around 1.33 eV. A noticeable difference between the two samples occurs in their polarization behavior: independent of the incident light polarization, sample B shows highly linearly polarized iX PL. While iX PL with a broad linewidth and peak intensity centered in the 1.32-1.35 eV range has been observed in other studies\cite{Rivera2015, Rivera2016, Miller2017, Jiang2018a, Jiang2018b}, the presence of linear polarization has not yet been reported. We have studied a total of eight MoSe$_2$/WSe$_2$ heterostructures showing linearly-polarized iX PL, and AFM investigations of four heterostructures with different emission types (two with circularly and two with linearly polarized iX PL) confirmed that the polarization behavior is directly linked to the sample morphology (see SI~Fig.~S2).

Linearly polarized luminescence arising from coherent superposition of intralayer excitons in the non-equivalent valleys has been previously observed in single-layer WSe$_2$\cite{Jones2013,Dufferwiel2018,Qiu2019} and WS$_2$\cite{Schmidt2016}, with the polarization axis defined by the polarization of the incident light, independent of the crystal orientation. In contrast, we find that iX PL linear polarization in sample B has an arbitrary and position-dependent orientation with respect to the incident light polarization, and is directly linked to the orientation of the sample. The inset in Figure 1e shows a polar plot of the integrated iX PL intensity as a function of the detected polarization angle measured at one of the locations in the MoSe$_2$/WSe$_2$ heterostructure region of sample B under linearly polarized excitation. For this measurement, the incident laser polarization was fixed at an angle indicated by the green arrow, and the angle of the detected linearly-polarized light was controlled using a linear polarizer and a rotating half-wave plate. Open black circles represent experimental data while the solid red line is a numerical fit using $I = A \times (1 + \rho \times \cos(2(\theta-\phi)))$, where $\theta$ is the detection angle, $\phi$ is the iX PL polarization angle, and A is a normalization constant. The detected iX PL signal has polarization axis  rotated by 50$^\circ$ in the counterclockwise direction with the respect to the laser polarization, with polarization degree $\rho = 0.28$. While sample A which shows excitation energy dependent circular polarization retention, the linear polarization degree in sample B appears to be unaffected by the excitation energy, remaining at the same level even when the latter is increased to 2.33 eV (see SI~Fig.~S3).

\begin{figure}[ht]
	\centering
	\includegraphics{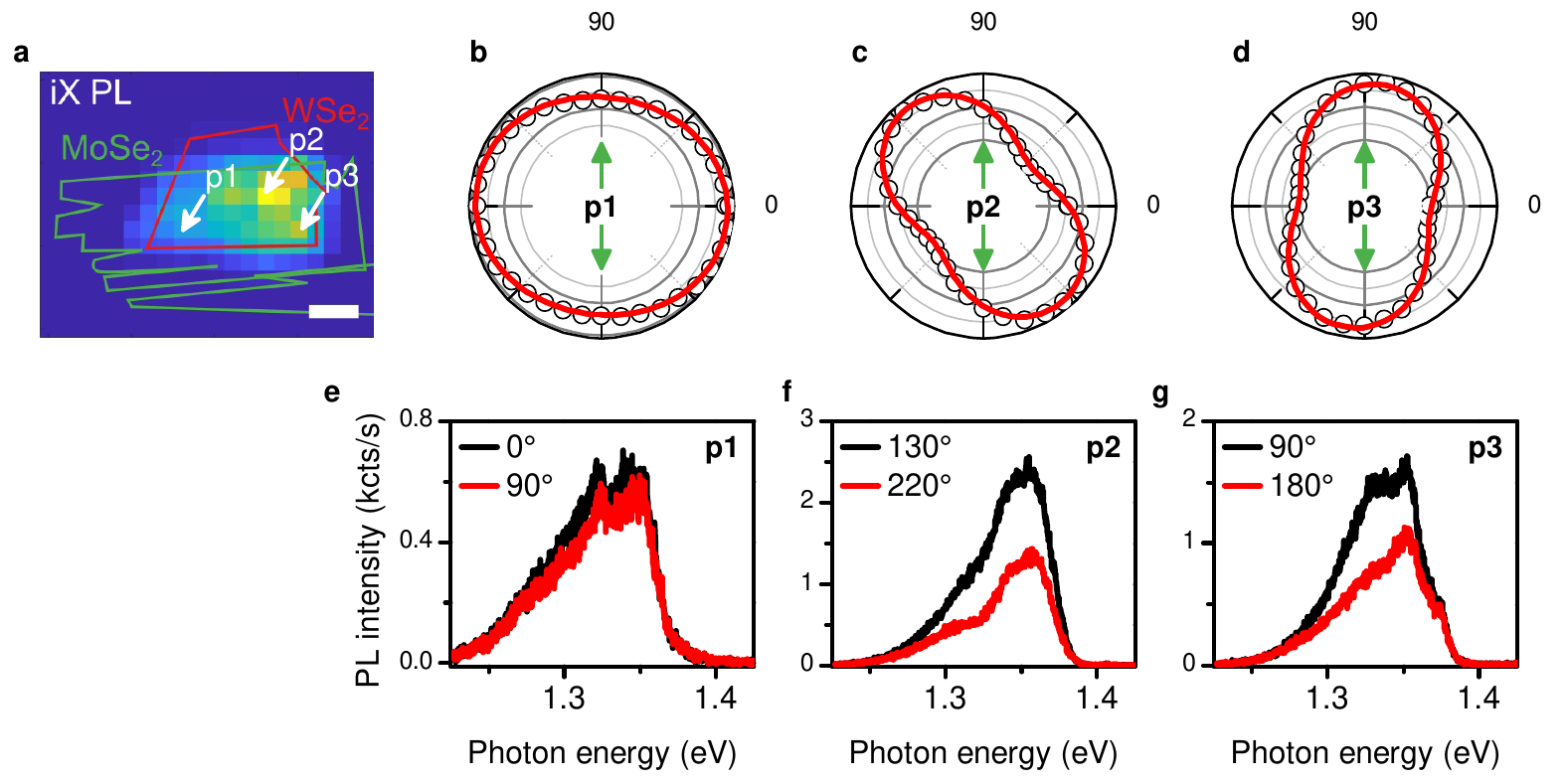}
	\caption{\textbf{Spatial variation of iX polarisation.} (a) Map of iX peak integrated intensity acquired in the sample shown in Fig.~\ref{fig:SampleImage}c. Solid lines indicate the edges of the MoSe$_2$ (green) and WSe$_2$ (red) monolayer regions. Scale bar, 3 um. (b), (c) and (d) Integrated intensity of the iX peak as a function of the detection angle recorded in regions P1, P2, and P3. (e), (f), and (g) polarization-resolved PL spectra recorded in the regions P1-P3 acquired at the detection angles corresponding to maximum (black) and minimum (red) integrated PL intensity.}
	\label{fig:SpatialVariation}
\end{figure}

Figure~\ref{fig:SpatialVariation}a plots a spatial map of the integrated iX PL intensity recorded in sample B, centered at the heterobilayer region. The low-energy emission in the 1.2-1.4~eV range is present only in the region where the two materials overlap, confirming its interlayer origin. The iX peak energy and shape, along with its linear polarization degree and orientation, demonstrate strong spatial variation, most likely stemming from the inhomogeneity of the transfer-induced strain in the sample.  Fig.~\ref{fig:SpatialVariation}b-d compares polar plots of integrated iX PL intensity collected in a similar way to the data in Fig.~\ref{fig:SampleImage} at the three points (P1-P3) of the heterostructure marked by white arrows in Fig.~\ref{fig:SpatialVariation}a. The PL spectra recorded at detection angles corresponding to maximum (black) and minimum (red) emission intensity are shown below each plot (Fig.~\ref{fig:SpatialVariation}e-g). The three points show distinctively different polarization properties: while P1 demonstrates emission with a very weak linear polarization aligned along the horizontal axis, the iX PL at P2 is highly linearly-polarized ($\rho \sim 0.29$), with the emission intensity maximum located at $\phi = 130^{\circ}$. On the other hand, P3, which is separated from P2 by approximately 2~$\mu$m, has the polarization axis rotated by 40$^{\circ}$ clockwise, with the polarization strength remaining at approximately the same level ($\rho \sim 0.27$). In addition to different polarization properties, the three points show a noticeable difference in the lineshape of the iX PL peak. As the typical distance between the nanowrinkles ($\sim$100 nm) is much smaller than the excitation spot size ($d \approx 1.5 \mu m$), the collected PL spectrum is a combination of signals coming from areas with different strain levels, resulting in the broad linewidth and the apparent multicomponent structure of the iX PL peak. Moreover, some regions show that different spectral components of the peak can have different polarization degrees and directions (SI~Fig.~S4). 
Note that, while the degree of linear polarization varies strongly from point to point, we have not observed noticeable circular polarization retention in any areas of sample B. In other samples investigated in this study we have occasionally detected the presence of both types of iX emission, most likely originating in the regions with different layer morphology (see SI~Fig.~S5).

\begin{figure}[ht]
	\centering
	\includegraphics{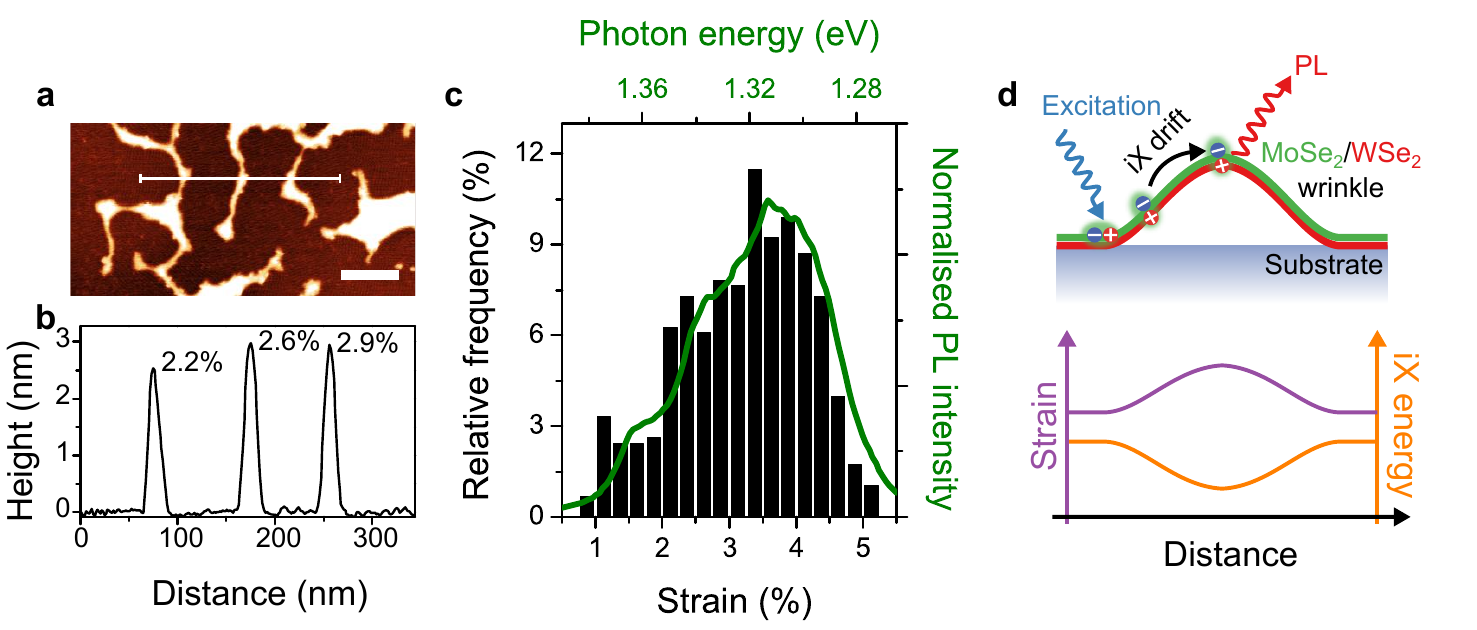}
	\caption{\textbf{Strain-induced interlayer exciton localization.} (a) Topographic AFM image of a single-layer WSe$_2$ deposited onto hBN. Black to white scale, 3.5 nm, scale bar, 100~nm. (b) Cross-sectional height profile measured from (a). Maximum values of tensile strain are listed above each wrinkle. (c) Histogram of maximum strain distribution in wrinkled WSe$_2$ monolayer extracted from an AFM image of a 1 $\mu$m $\times$ 0.2 $\mu$m region. The green curve plots an estimated PL spectrum of an ensemble of wrinkle-trapped iXs, calculated as a sum of emission originating from regions with levels of strain. (d) Schematic diagram illustrating drift and accumulation of iX at transfer-induced nanowrinkles caused by the strain-induced band gap reduction.}
	\label{fig:iXStrain}
\end{figure}

The observation of circularly-polarized iX PL in MoSe$_2$-WSe$_2$ heterostructures with flat surface morphology is a direct consequence of the three-fold rotational symmetry of the crystal lattices of the constituent layers \cite{Wang2017a,Rivera2018}. Negligible circular polarization retention and the emergence of linearly-polarized PL in wrinkled samples provides strong evidence of symmetry breaking. We attribute this change of optical properties to iX accumulation at local strain maxima, located at the top of the transfer-induced nanowrinkles. The relaxation of strain induced in the sample by the stretching and compression of the polymer carrier film during the transfer occurs through buckling of the flake. This results in the formation of nanowrinkles that accumulate stress, separated by flat strain-free regions\cite{Vella2009,Castellanos-Gomez2013}. A typical height image of transfer-induced nanowrinkles in monolayer WSe$_2$ and its cross-section are shown in Figure~\ref{fig:iXStrain}a and b. The maximum tensile strain found at the top of the nanowrinkle can be estimated as $\varepsilon \sim \pi^2 h \delta/((1-\sigma^2)\lambda^2)$, where h is the thickness of the flake, $\delta$ is the height and $\lambda$ is the width of the wrinkle, and $\sigma$ is the Poisson's ratio\cite{Castellanos-Gomez2013}. The histogram in Fig.~\ref{fig:iXStrain}c plots a typical distribution of strain in corrugated WSe$_2$, extracted from Fig.~\ref{fig:SampleImage}e by measuring cross-sectional profiles of individual wrinkles. The distribution has an approximately symmetric shape, with the mean value of 3.2\% and standard deviation of 1.0\%. While in the samples that we studied surface deformation occurs primarily in the first transferred layer, similar width of the nanowrinkles in monolayer and heterobilayer regions indicate that the top layer conforms to the morphology of the underlying material, likely resulting in a similar level of strain. 

As the emission energy of iXs is predominantly defined by the band offset between the two materials (Fig.~\ref{fig:SampleImage}a), the presence of tensile strain can lead to a red-shift of the iX peak, which occurs due to the strain-induced reduction of the interlayer band gap. Indeed, red-shift of iX emission energy has been observed in strained MoSe$_2$/WSe$_2$ samples at room temperature, with the slope of 22.7 meV/\%\cite{He2016}. Strain-induced reduction of the interlayer band gap can lead to funnelling of iX towards the local strain maximum, located at the top of the nanowrinkles (Fig.~\ref{fig:iXStrain}d). With typical lifetime in the nanosecond range, iXs can diffuse over large distances\cite{Jauregui2018} and accumulate in the corrugated areas, resulting in the PL originating predominantly from these regions, as schematically shown in Fig.~\ref{fig:iXStrain}d. We calculate an estimated emission spectrum of wrinkle-trapped iXs as a sum of Lorentzian peaks with the parameters typical for high-quality samples (emission energy of 1.4 eV and FWHM of 10 meV), red-shifted from their position in unstrained heterostructures according to their strain level, with the relative intensity defined by the corresponding bin counts. This simple model produces a surprisingly good reconstruction of the iX PL spectrum: the simulated spectrum (green curve in Fig.~\ref{fig:iXStrain}c) reproduces the apparent multi-peak structure, with a FWHM of 60~meV and an intensity maximum located at $\approx$1.32~eV.  

\begin{figure}[ht]
	\centering
	\includegraphics{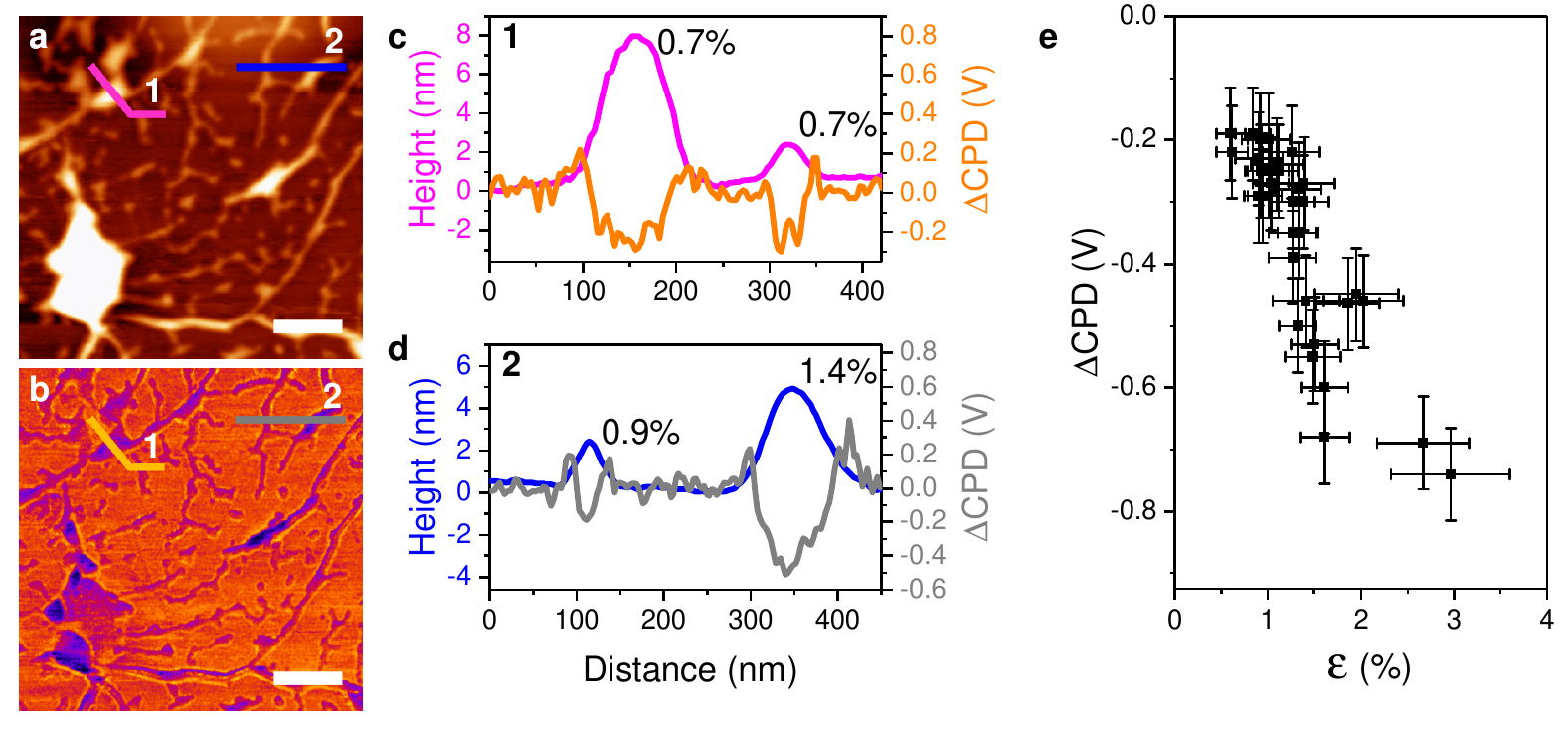}
	\caption{\textbf{Surface potential variation in wrinkled heterostructures.} (a) Topographic AFM image of a wrinkled MoSe$_2$/WSe$_2$ heterostructure region recorded on sample B. Black to white scale, 9 nm, scale bar, 300~nm. (b) KPFM image of the same region showing lower contact potential difference (CPD) in wrinkled regions, indicating an increase in the local work function. Scale bar, 300 nm; color scale, 1.5V. (c and d) Cross-sectional profiles extracted from AFM and KPFM images of the regions marked with corresponding colored lines in (a) and (b). Numbers above the curves indicate the value of tensile strain at the top of the wrinkle estimated from wrinkle topography. (e) CPD offset as a function of maximum tensile strain extracted from AFM and KPFM images.}
	\label{fig:KPFM}
\end{figure}

In order to confirm that the transfer-induced wrinkles can act as efficient iX accumulation centers, we have characterized the potential landscape of the sample using KPFM. This technique allows the local work function to be mapped by measuring the contact potential difference (CPD) between a conductive AFM tip and the sample surface\cite{Melitz2011} and has been successfully applied for 2D materials characterization\cite{Li2013,Li2015d,Yu2009}. Figure~\ref{fig:KPFM}a and b compares AFM and KPFM images recorded in a heterobilayer area of sample B. The variation of the CPD shows a clear correlation with topographical features: while regions with flat layer morphology demonstrate a homogeneous signal, a strong variation of the CPD can be observed in the transfer-induced nanowrinkles. Fig.~\ref{fig:KPFM}c and d plot cross-sectional profiles of the topographic and surface potential images extracted from the regions marked by the colored lines in Fig.~\ref{fig:KPFM}a and b. All four nanowrinkles demonstrate similar potential profiles: compared to the flat heterostructure areas, the bases of the wrinkles show a positive CPD contrast, which becomes negative at the center of the wrinkle, with the strongest contrast observed at the highest point of the wrinkle. The maximum contrast value is not defined by the wrinkle amplitude, with similar values observed in the wrinkles with substantial height difference (Fig.~\ref{fig:KPFM}c). However, it is strongly correlated with the tensile strain value extracted from the wrinkle height profile. The CPD signal measured in KPFM is a surface potential variation that reflects the difference in the work function between the conductive tip and the sample surface. This variation can have many causes\cite{SaschaSadewasser2011}, including local band bending, change in the carrier concentration, and a change in the surface chemistry, for example due to the presence of contamination on the sample surface. In order to confirm that the observed surface potential variation has strain-related origin, we have compared CPD values in wrinkles having different strain levels, as estimated from AFM measurements. Fig.~\ref{fig:KPFM}e plots the maximum value of the CPD offset as a function of tensile strain estimated from the wrinkle topography, which was recorded simultaneously with the KPFM data. The two parameters show a clear correlation, with higher values of tensile strain leading to stronger negative offset of the CPD, providing further evidence of strain as the primary source of the observed surface potential variation. The overall lower strain level compared to that extracted from the AFM data in Fig.~\ref{fig:iXStrain} is likely a result of a difference in the probes used for the two measurements. While probes with the nominal tip radius of 10~nm were used for the AFM imaging, the KPFM measurements were carried out using a conductive probe with 25 nm nominal radius. In the latter case, the larger size of the tip limited the spatial resolution of the acquired topographic images, resulting in a lower measured wrinkle aspect ratio and consequently a lower extracted strain level. 

The spatial variation of the surface potential observed in our measurements is in agreement with the predicted strain-induced band gap variation. Indeed, compressive strain and consequent band gap widening are predicted to arise in the regions with the concave (downward-bent) profile, which can be found at the bases of the wrinkles\cite{Brooks2018}, where we found a positive KPFM contrast (Fig.~\ref{fig:KPFM}c and d). Conversely, the convex regions of the wrinkles found at their tops are expected to show the reduction of the band gap caused by tensile strain, consistent with the observed decrease in the KPFM contrast. Local modification of band gap induced by strain leads to efficient funnelling of iXs from flat unstrained regions towards local energy minima located at the top of the nanowrinkles, resulting in the PL signal dominated by wrinkle-trapped iX emission. The potential barriers formed due to the compressive strain present at the bases of the wrinkles can further enhance iX trapping efficiency, although their characteristics are defined by wrinkle geometry and vary substantially between different wrinkles. As the reduction of the interlayer band gap is defined by the tensile strain level, the local energy minima will coincide with regions with the highest crystal anisotropy, resulting in the emergence of linearly-polarized iX PL, with the orientation of the polarization axis defined by the wrinkle geometry. The polarization of the detected iX PL signal will therefore be defined by the density, dimensions, and orientation of the nanowrinkles within the excitation spot. Regions with highly polarized iX PL likely correspond to areas with higher levels of strain and/or a preferential nanowrinkle alignment.

In conclusion, we have investigated the emergence of linearly-polarized interlayer exciton photoluminescence in MoSe$_2$/WSe$_2$ heterostructures with a high density of nanometer-scale wrinkles which form during the monolayer transfer process. Using a combination of optical spectroscopy and atomic force microscopy measurements, we have traced its origin to the accumulation of the interlayer excitons at the top of the transfer-induced nanowrinkles, where strong tensile strain leads to the reduction of the interlayer band gap, as well as breaking of the local crystal symmetry. Surface potential images of the sample acquired using Kelvin probe force microscopy reveal variations of the work function consistent with the strain-induced band modulation, with the maximum contact potential difference showing a strong correlation with the strain level extracted from the nanowrinkle shape. 

The presence of trapped contamination and residual strain are issues inherent to van der Waals heterostructure fabrication and often is the main factor limiting the performance of devices manufactured using mechanical transfer. Considerable efforts have been dedicated to the optimization of fabrication and processing methods in order to reduce the disorder arising from charged impurities, however, the problem of transfer-induced strain anisotropy has been largely unaddressed. While the presence of random residual strain is detrimental to the device optical quality, controlled creation of wrinkles in the heterostructures can be achieved using pre-patterned substrates\cite{ Kremser2019} or by utilizing the difference in thermal expansion coefficients \cite{Oliveira2015}. The resulting highly localized strain fields can be utilized to control localization and propagation of interlayer excitons, opening interesting possibilities for designing novel micro-optical circuits based on van der Waals heterostructures. 

\subsection{Methods}
\textbf{Photoluminescence spectroscopy.}
Spectrally resolved PL measurements were performed in a custom-built micro-PL setup. Excitation light centered at 1.88~eV (660~nm) generated by a diode laser (ADL-66505TL, Roithner) was focused onto the sample using a 50x objective lens (M Plan Apo 50X, Mitutoyo). For excitation-energy dependent measurements, a 532 nm diode-pumped solid-state laser (CW532-050, Roithner) and a 730 nm diode laser (HL7302MG, Thorlabs) were used. The PL signal collected in the backwards direction was isolated using an appropriate shortpass filter and detected by an 0.5~m spectrometer (SP-2-500i, Princeton Instruments) with a nitrogen cooled CCD camera (PyLoN:100BR, Princeton Instruments). The low temperature measurements were carried out using a continuous flow liquid helium cryostat, where the sample was placed on a cold finger with a base temperature of 10~K.

\textbf{Sample fabrication.}
MoSe$_2$/WSe$_2$ heterostructures were assembled using a standard dry-peel transfer technique. Monolayer MoSe$_2$ and WSe$_2$ flakes were mechanically exfoliated from bulk crystals (provided by HQ Graphene) onto a silicon substrate coated with a polymer bilayer composed of polymethylglutarimide (PMGI) and polymethyl methacrylate (PMMA). After identifying monolayer flakes using optical contrast measurements and photoluminescence spectroscopy, the bottom PMGI layer was selectively dissolved using a water-based developer (MF319, Microposit), releasing the 1~$\mu$m thick PMMA membrane from the substrate. Vertical heterostructures were assembled by consecutively transferring TMD flakes onto a thin hBN crystal that was mechanically exfoliated onto a SiO$_2$/Si substrate. The PMMA carrier film was then removed by mechanical peel off.

\textbf{Atomic force microscopy.} AFM images of the samples were acquired in tapping mode using a Dimension Fastscan (Bruker) and Arrow UHFAuD cantilevers (NanoWorld). The cantilever was re-tuned close to the surface and the setpoint, feedback gains and Z range were optimized during imaging. 


\textbf{Kelvin probe force microscopy.} 
KPFM measurements were carried out with a Nanotec Electronica AFM\cite{Horcas2007,Gimeno2015} using PtIr-coated PPP-EFM probes (Nanosensors), with nominal stiffness and resonance frequency of 2.8 Nm~$^{-1}$ and 75~kHz, respectively. Simultaneous dynamic mode topography and frequency modulation mode KPFM in a single-pass scheme were used\cite{SaschaSadewasser2011}.For the topography, an amplitude set point of 25~nm (cantilever free amplitude 35~nm) was used, whereas for the KPFM an alternate bias voltage of amplitude 3~V and frequency 7~kHz was employed.

\begin{suppinfo}
Large-area AFM scans of layer morphology in mechanically stacked MoSe$_2$/WSe$_2$ heterostructures; low-temperature PL spectra and AFM images of samples with different surface morphology; excitation energy dependence  of polarization degree in samples with circularly and linearly polarized iX emission; polarization angle of iX peak components;  co-existence of different iX PL types. 
\end{suppinfo}

\begin{acknowledgement}
E. M. A. and A. I. T. thank the financial support of the Graphene Flagship under grant agreements 696656 and 785219, and EPSRC grants EP/P026850/1, EP/S030751/1 and EP/N031776/1. P. A., H. N-A., and L. F. received funding from the Marie Sklodowska-Curie Actions (grant agreement No 793394) and the European Research Council (grant agreement No. 819417) under the European Union’s Horizon 2020 research and innovation programme. Authors thank F. Guinea and M. Brooks for fruitful discussions.

\end{acknowledgement}



\providecommand{\latin}[1]{#1}
\makeatletter
\providecommand{\doi}
  {\begingroup\let\do\@makeother\dospecials
  \catcode`\{=1 \catcode`\}=2 \doi@aux}
\providecommand{\doi@aux}[1]{\endgroup\texttt{#1}}
\makeatother
\providecommand*\mcitethebibliography{\thebibliography}
\csname @ifundefined\endcsname{endmcitethebibliography}
  {\let\endmcitethebibliography\endthebibliography}{}


\end{document}



\newpage
\renewcommand{\thefigure}{S\arabic{figure}}%

\begin{figure}[h]
	\centering
	\includegraphics{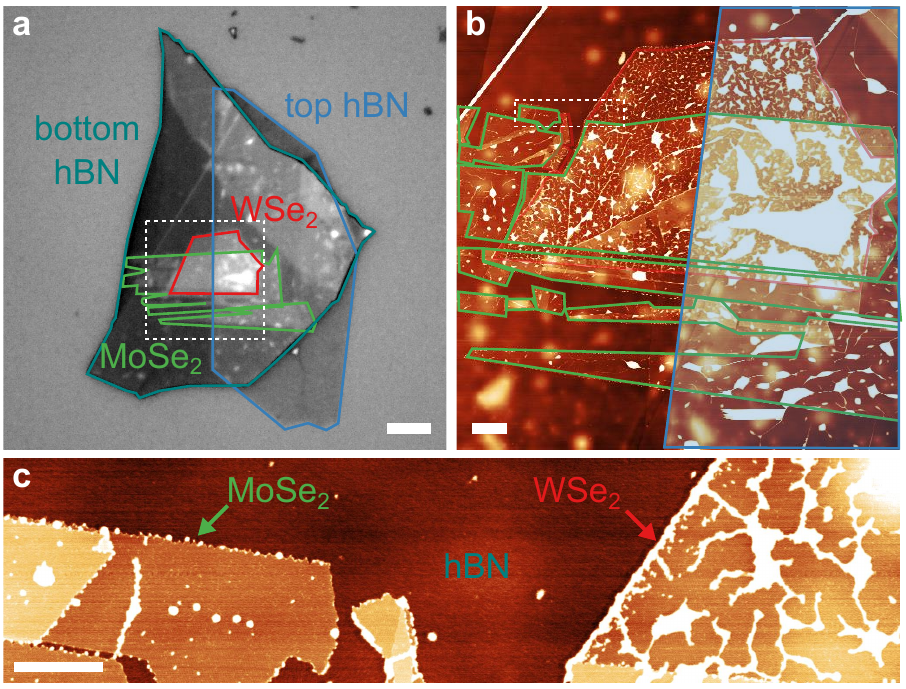}
	\caption{\textbf{Layer morphology of mechanically transferred TMD flakes.} (a) Optical image of a MoSe$_2$/WSe$_2$ heterostructure assembled on a hBN/SiO$_2$/Si substrate using dry-peel mechanical transfer. A region of a heterostructure was subsequently encapsulated with a thin hBN layer. Scale bar, 5~$\mu$m. (b) AFM height image recorded in the region of the sample indicated by the dashed square in (a). Edges of MoSe$_2$, WSe$_2$, and hBN layers are marked by green, red, and blue lines, respectively. Strong layer corrugation is visible in the WSe$_2$ monolayer and MoSe$_2$/WSe$_2$ heterobilayer regions, but is absent in the monolayer MoSe$_2$ flakes and the underlying hBN. Scale bar, 1~$\mu$m; color scale, 10~nm. (c) AFM height image showing single-layer MoSe$_2$ and WSe$_2$ regions (recorded in the area indicated by the dashed rectangle in (b)). Scale bar, 300~nm; color scale, 3~nm.}
	\label{fig:SI_AFM}
\end{figure}

\begin{figure}[h]
	\centering
	\includegraphics{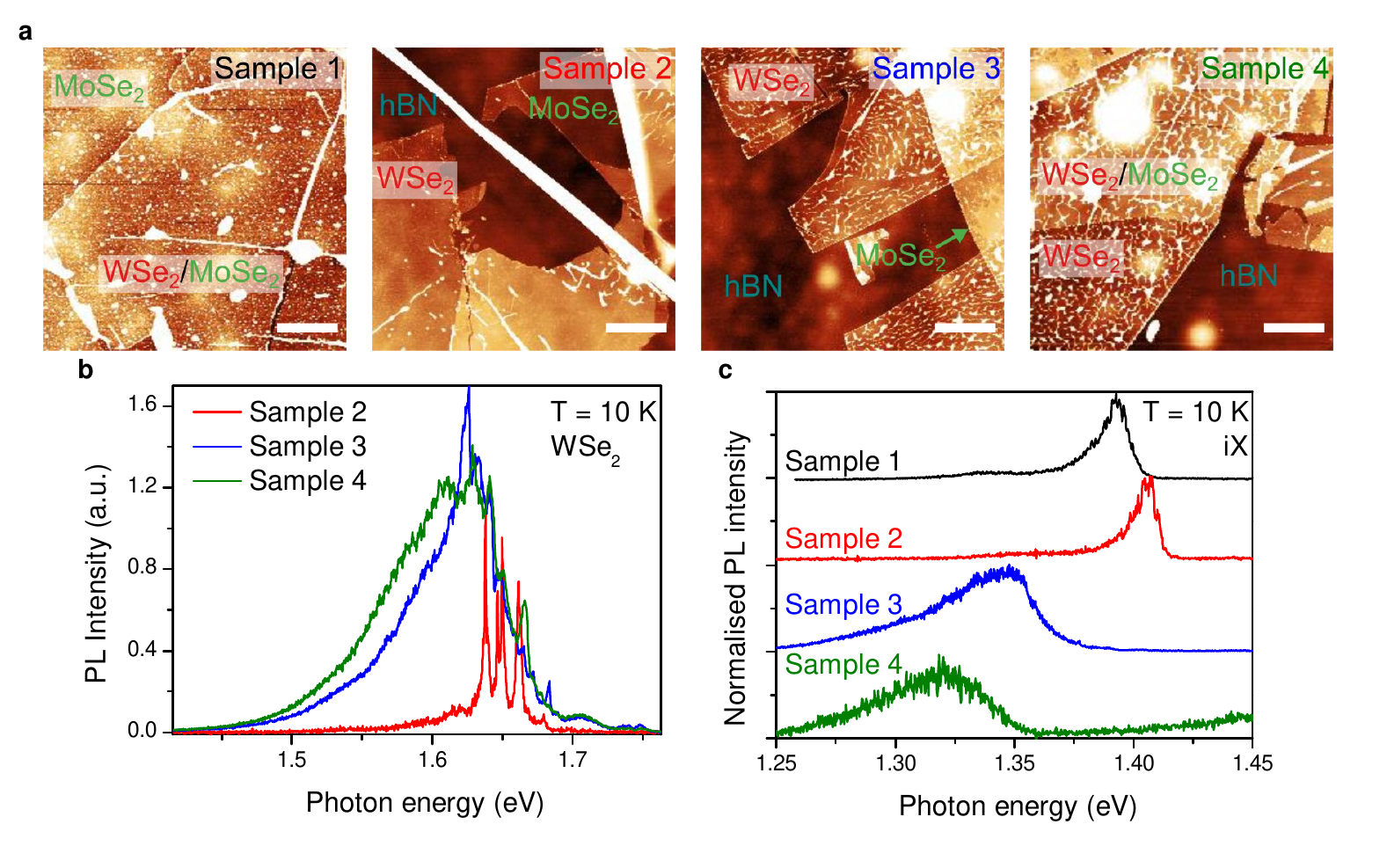}
	\caption{\textbf{Optical properties of samples with different layer morphology.} (a) AFM height images of MoSe$_2$/WSe$_2$ heterostructures with flat (Samples 1 and 2) and wrinkled (Samples 3 and 4) layer morphology. Scale bars, 1~$\mu$m; color scale, 4~nm.(b) Photoluminescence spectra recorded at T~=~10~K in single-layer WSe$_2$ regions of Samples 2, 3, and 4. Samples 3 and 4 (with strong layer corrugation) show substantially broader exciton PL lines and strong low-energy defect emission band. PL spectrum in Sample 1 is not shown as the sample did not contain an isolated WSe$_2$ monolayer area. (c) Low-temperature PL spectra recorded in samples 1-4, centered at the interlayer exciton peak.}
	\label{fig:SI_iX_types}
\end{figure}

\begin{figure}[h]
	\centering
	\includegraphics{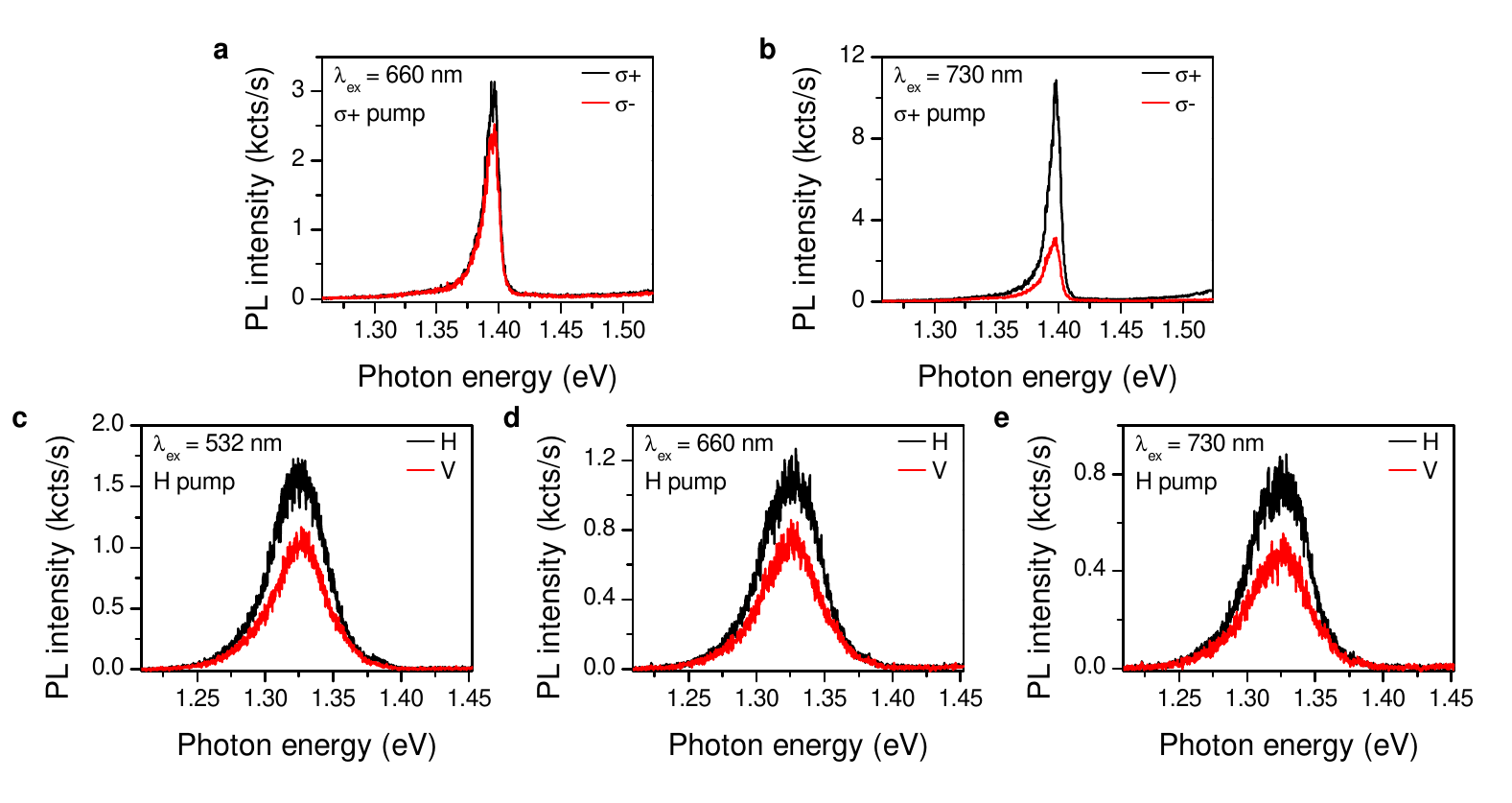}
	\caption{\textbf{Excitation energy dependence of the iX PL polarization.} (a) and (b) Helicity-resolved iX PL spectra recorded on a MoSe$_2$/WSe$_2$ sample with flat layer morphology at T = 10 K using (a) 660~nm and (b) 730~nm laser excitation. Black and red lines correspond to co- and cross-polarized detection, respectively. (c-d) iX PL spectra recorded on a wrinkled MoSe$_2$/WSe$_2$ sample under linearly-polarized excitation aligned along the horizontal axis and centered at (c) 532 nm, (d) 660 nm, and (e) 730 nm. Black (red) lines correspond to the signals recorded at the detection angle parallel (perpendicular) to the incident light polarization. }
	\label{fig:SI_exWavelength}
\end{figure}

\begin{figure}[h]
	\centering
	\includegraphics{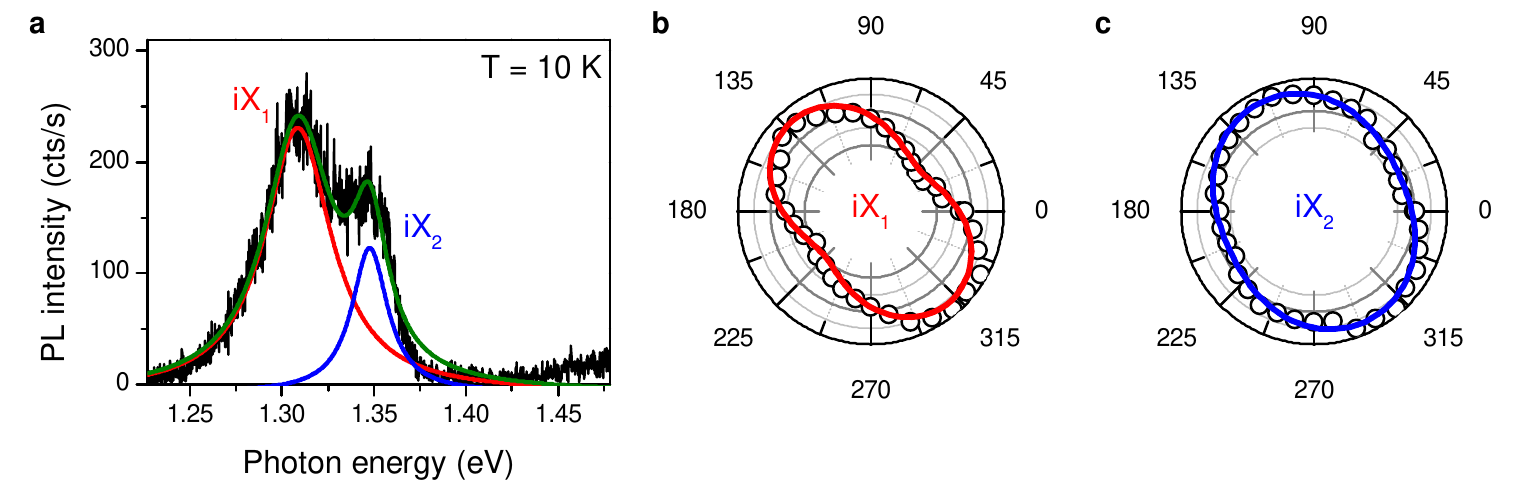}
	\caption{\textbf{Polarization axis direction of iX peak components.} (a) iX PL spectrum recorded on a wrinkled MoSe$_2$/WSe$_2$ sample showing two peak components iX$_1$ and iX$_2$ with different polarization direction. Red and blue lines correspond to Lorentzian fits to the data. (b) and (c) Integrated intensity of iX$_1$ and iX$_2$ peaks as a function of detection angle. iX$_1$ shows the maximum PL intensity at $\phi = 130^\circ$, with polarization degree $\rho = 0.29$. The iX$_2$ demonstrates a weaker polarization degree $\rho = 0.15$, with the PL maximum at $\phi = 115^\circ$.}
	\label{fig:SI_iXcomponents}
\end{figure}

\begin{figure}[h]
	\centering
	\includegraphics{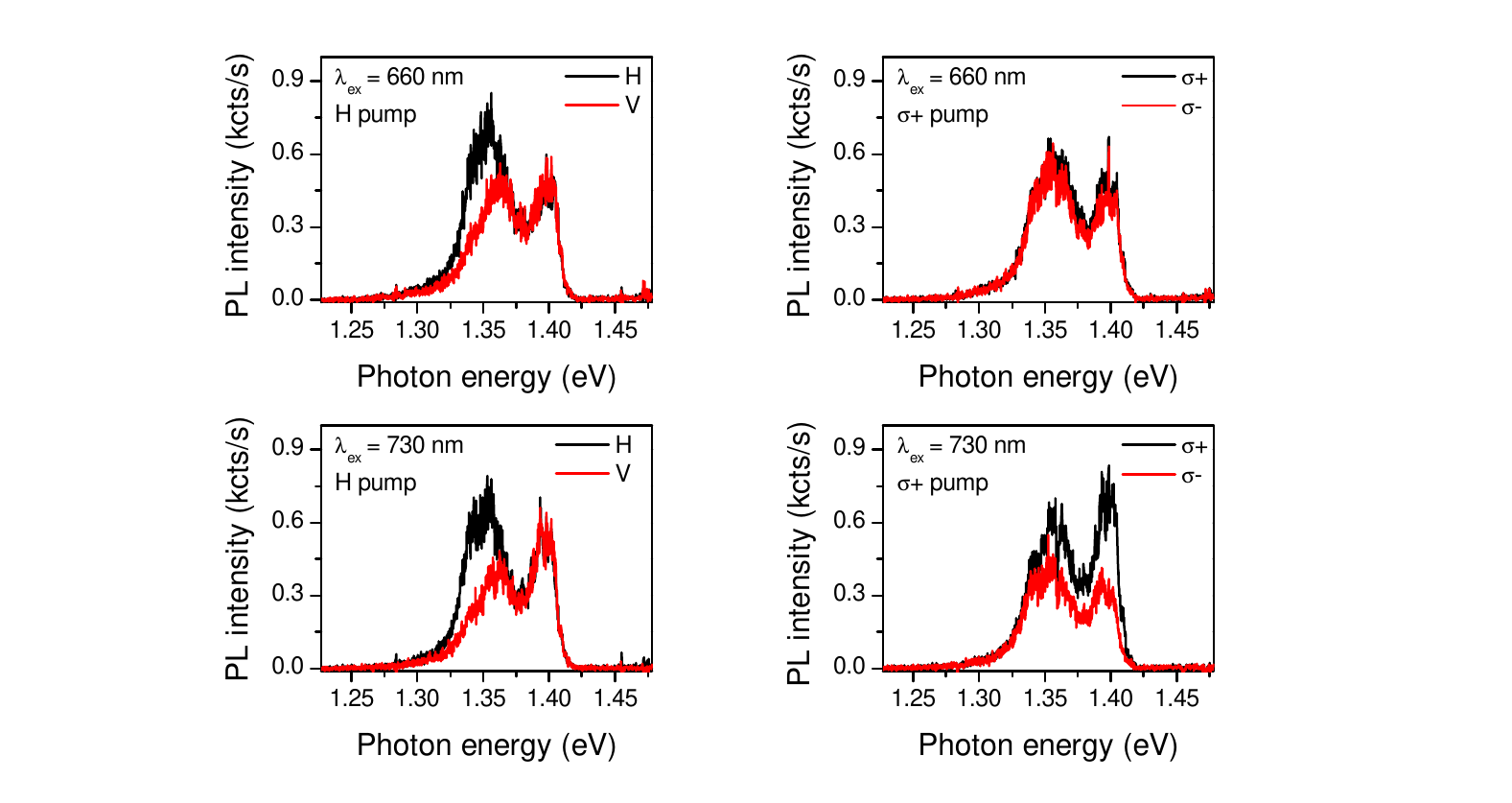}
	\caption{\textbf{Co-existence of different iX PL types.} iX PL spectra recorded at T = 10 K under different excitation conditions showing two iX peaks with different polarization behavior; excitation laser parameters are listed in each panel. The higher-energy component shows a negligible degree of linear polarization under excitation with both 660 and 730 nm linearly-polarized laser light. Under circularly-polarized excitation centered at 730 nm, it demonstrates circular polarization retention with $\rho_c = 0.32$, however, circular polarization degree becomes negligible under 660 nm excitation, consistent with the iX behavior observed in samples with flat surfaces. In contrast, lower-energy peaks show linear polarization degree $\rho_l = 0.27$ under excitation with both 660 and 730 nm light. Note that the linear polarization degree falsely appears lower under circularly-polarized light excitation due to the quarter-wave plate installed in the beam path that transforms it into an elliptically-polarized signal. }
	\label{fig:SI_C+Lemission}
\end{figure}